# Towards scalable nano-engineering of graphene

A.J. Martínez-Galera[1], I. Brihuega[1,2]*, A. Gutiérrez-Rubio[1,3], T. Stauber[1,2,3], J. M. Gómez-Rodríguez[1,2].

AUTHOR ADDRESS.

[1]Departamento Física de la Materia Condensada, Universidad Autónoma de Madrid, E-28049 Madrid, Spain.

[2]Condensed Matter Physics Center (IFIMAC), Universidad Autónoma de Madrid, E-28049 Madrid, Spain.

[3]Instituto de Ciencia de Materiales de Madrid, Consejo Superior de Investigaciones Científicas, E-28049 Madrid, Spain.

*) Corresponding author: ivan.brihuega@uam.es

**By merging bottom-up and top-down strategies we tailor graphene's electronic properties within nanometer accuracy, which opens up the possibility to design optical and plasmonic circuitries at will. In a first step, graphene electronic properties are macroscopically modified exploiting the periodic potential generated by the self assembly of metal cluster superlattices on a graphene/Ir(111) surface. We then demonstrate that individual metal clusters can be selectively removed by a STM tip with perfect reproducibility and that the structures so created are stable even at room temperature. This enables one to nanopattern circuits down to the 2.5 nm only limited by the periodicity of the Moiré-pattern, i.e., by the distance between neighbouring clusters, and different electronic and optical properties should prevail in the covered and uncovered regions. The method can be carried out on micro-meter-sized regions with clusters of different materials permitting to tune the strength of the periodic potential.**

Two main routes are usually followed to modify graphene's electronic and optical properties. On the one hand, bottom up approaches have proven to be efficient to change the overall electronic structure of graphene, enabling for example, the gap opening at the Fermi energy [1-3], renormalization of the Fermi velocity [4-6] or controllable n- and p-type electronic doping [7-10]. On the other hand, with top down approaches it is possible to induce these alterations on a local scale enabling one to pattern graphene to quantum confine electrons [11-14], to induce local magnetic and superconducting properties[15,16], or to use a scanning probe to selectively tune its electronic properties [13,17,18]. Still, a remaining challenge is the realization of controlled nanopatterning below 10nm sizes [19,20], key for the comprehensive integration of graphene in real devices. Here, we show that combining both approaches, i.e., bottom-up and top-down, one can reach a 2.5nm patterning, enriching graphene's capabilities even more.

Let us first outline the bottom-up approach[1,2,4,5,21,22] for graphene monolayers on several metallic substrates which can be epitaxially grown with unrivaled quality[23]. An interesting common feature of most of these graphene-metal interfaces is the presence of superperiodicities, known as Moiré patterns, resulting from the lattice mismatch and rotation angle between graphene and metal lattices [24]. This creates a periodic potential superimposed to graphene whose strength can be tuned by the preferential adsorption of different adsorbates on specific positions of the Moiré superlattice [1-3,25]. A graphene metal interface particularly interesting for our purposes is the graphene monolayer epitaxially grown on Ir(111) substrates. It allows for growing single Moiré domains extending over micrometers[25,26] while at the same time, the interaction with the substrate remains weak leaving almost unaltered the electronic properties of the graphene layer, i.e., the $\pi$-bands with the characteristic linear dispersion and Fermi velocity of free standing graphene are only modified by the appearance of a small gap less than 100 meV,[2,27,28] see Fig.1b.

Additionally, this Moiré pattern formed by the graphene monolayer and the Ir(111) substrate can be used as a template for networks of monodisperse clusters of transition metals.[25,29] As recently reported, the adsorption of these cluster superlattices strengthens the periodic potential created by the Moiré pattern, modifying the electronic properties of the graphene layer.[2] In particular, an increase of the band gap up to 400 meV and large anisotropies of the electron group velocity close to the Dirac point have been measured for Ir cluster superlattices[2], see Fig 1c.

**RESULTS:**

The experimental bottom-up procedure is the following. We first grow a graphene monolayer on an Ir(111) substrate by chemical vapor deposition (CVD) of ethylene in UHV environments with the Ir(111) substrate held at 1050ºC. Then, by evaporationg W or Ir from high purity filaments, we subsequently cover it with a hexagonal array of metal clusters with 2.5nm periodicity (see methods for details on the sample preparation).

We will now turn to the above mentioned top-down approach where STM appears as an ideal technique to tackle local manipulation on such samples due to its ability to modify and pattern 2D samples with ultimate resolution [30-32]. In Fig 1a, we write "graphene" on the G/Ir(111) surface using the STM tip to completely remove the selected clusters which demonstrates the patterning on top of graphene with 2.5nm accuracy and a very high degree of complexity, see also examples of Figs 1 e-g. In this way, by deliberately removing metallic clusters from the graphene layer, we can recover the electronic properties corresponding to the pristine G/Ir(111) interface in specific regions of the sample and can architecture nanostructures formed by two different kinds of *'graphene'* regions, i.e., ones covered with clusters and uncovered ones, with supposedly different electronic properties. It is noteworthy that our method can be used with clusters of different metallic elements, see Figs 1e-g where Ir (e-f) and W (g) clusters formed by approx. 50 atoms have been removed. This might allow tuning the strength of the periodic potential superimposed to the graphene layer and, consequently, the electronic and optical properties for the covered regions.

The procedure we have developed to engineer graphene nanostuctures consists in selectively removing single metallic clusters on top of graphene by gently approaching the STM tip towards them, as schematized in Fig 2a. We first image a large graphene sample area completely covered with metallic clusters, see Fig 2b. Next, we choose a metal cluster to be removed and stop the STM tip above it. With the tip above the chosen cluster, we open the feedback loop and bring the tip towards the sample at a constant rate for a distance of typically 0.6 nm. Then, we retract the tip back and close the feedback loop returning to the initial tunneling conditions. This completely removes the selected cluster as shown in Fig 2c. Finally, we systematically repeat this procedure to remove all selected clusters and thus form the designed nanostructure. As an example, the complete sequence for writing a "C" by consecutively removing 9 Ir clusters is shown in Figs 2b-k

During the patterning process, the tip resolution is very robust and we usually observe almost no changes in our resolution after each cluster removal (see supplementary material). It further appears that the extracted clusters wet the STM tip and indeed, we remove the metal cluster as a whole since no traces of metal atoms are observed on the graphene surface after each single extraction event. This is likely due to the large cohesive energy of both W and Ir compared to the binding energies of C-Ir and C-W, respectively.[25,33] Such high cohesive energies together with the strong W-Ir binding should thus be responsible of the observed tip stability; once the cluster material wets the STM tip, it remains there in an extremely stable manner such that we have not been able to place the metal cluster back on the graphene surface.

The possibility of picking up or manipulating individual clusters formed on the Moiré-pattern was previously mentioned[25,34]. But on these works the cluster manipulation was a rather rare and statistical event. In fact, it was even considered a disturbing effect since it happened more or less statistically during the scanning process that could only be avoided under suitable tunneling conditions. Our work thus goes far beyond these earlier observations as we are now able to demonstrate that these cluster manipulations can be controlled to form arbitrary patterns stable even at room temperature. An essential issue regarding the validity of the procedure just described thus stems from its actual efficiency to extract the selected clusters.

To this end, we have performed a careful study of the probability of removing a cluster as a function of both the tip-sample approaching distance and the bias voltage applied to the sample during the whole process, see Fig. 2l and the supplementary material for details. The most important finding is that, for all voltages investigated during this study, we can reach a 100% probability for extracting a cluster by approaching the tip towards the sample a distance exceeding a certain value, between 0.5-0.7 nm, slightly different for each voltage. This allows to nanopattern the graphene surface with almost any degree of complexity and perfection. We also observed that, while the probability of extracting a cluster strongly depends on the approaching distance, the dependence on the applied voltage is much more moderate and basically independent of the voltage polarity, i.e., the direction of the electric field between tip and sample. For all voltages investigated here, the shape of the probability curves is essentially the same with only a rigid shift between them. This shift originates from the initial tip-sample distance dependence on the bias voltage set prior to open the feedback loop. Thus, our results point to a cluster removal procedure mainly driven by the actual distance between the STM tip and metal cluster.

To get more insight into the physical processes involved in the cluster extraction, we recorded the current during the vertical displacement of the STM tip (I-Z curves), see for example inset of Fig 2m. As usual when investigating the approach between two metallic electrodes, individual conductance curves were inherently irreproducible (see supplementary material), which is generally attributed to variations in the actual atomic-scale configuration of the metallic electrodes during the transition from tunneling to direct contact [35-38]. Thus, to perform an objective analysis of our experimental data, we constructed a conductance histogram from the evolution of the conductance traces of more than a thousand single cluster extraction events, see Fig 2m. Peaks in such conductance histograms are related to statistically more probable configurations in the contact formation [36-38]. The histogram shows a clear peak for a quantum of conductance ($G_0 = 2e^2/h$, $e$: electron charge; $h$: Planck's constant), indicating that the extraction of a metal cluster involves the formation of an atomic size contact. Similar $G \approx G_0$ values have been reported for contacts between an atomically sharp Au tip and graphene regions strongly bonded with a metal substrate[39]. In such regions, carbon atoms bind strongly to the metal surface and the hybridization of the graphene orbitals is transformed from $sp^2$ to $sp^3$, in a similar way as reported for graphene regions underneath metal clusters on the graphene/Ir(111) system [33].

Let us now address several key points to infer the actual potential of our method to architecture functional graphene nanostructures, in particular, size limits, stability and quality. The range of applicability is obviously limited by the size of the nanostructures that can be created. We can build nanostructures from the 2.5 nm limit given by the Moiré pattern distance to the few micrometers one which is given by the typical STM scanning range; the possibility of growing single Moiré patterns domains extends over several micrometers [26]. As an example, a 0.25 x0.25 µm$^2$ STM image of a graphene region uniformly covered by metallic clusters is shown in Fig 3a. Since the graphene layer grows as a carpet on top of the Ir substrate [40], monoatomic steps, as the one existing in the middle of the image, have very little influence on the cluster superlattice. Another important question deals with the stability of the created nanostructures since any practical application would require them to be stable at room temperature. Previous studies found that the cluster superlattices as a whole are stable up to temperatures of 400 K [29]. Here, we investigated the room temperature stability of several nanostructures with very different shapes and found them to be perfectly stable within our time scale (days). As an example, we show in Figs 3b-c two STM images acquired with 24 hours difference on the same sample area where an "A" nanostructure constructed by removing 10 W clusters and presenting a single isolated cluster in its center can be appreciated. The comparison of both images clearly reveals that even complex nanostructures keep exactly the same appearance one day after their construction.

Finally, we want to comment on the state of the graphene layer after the removal of the clusters. We aim to use pristine graphene on Ir(111) domains as one of our building blocks, thus, we need our cluster extraction method to produce perfectly clean graphene regions. To this end, we show in Figs 3d-f a sequence of STM images illustrating the evolution of a region where we have removed a large number of clusters. First, we show a STM image with the pristine W cluster superlattice, see Fig. 3d. Then, in Fig 3e, we show exactly the same sample region after using the STM tip to remove all the metal clusters from its central part. Last, in Fig 3f, we show an atomically resolved STM image of the cleaned region, acquired in the central area outlined by a blue square in Fig 3e. As can be observed, no single trace of the metal clusters is found on the cleaned region which is indistinguishable from the ones obtained on pristine graphene on Ir(111) prior to the W cluster.

**DISCUSION:**

The full potential and applicability of our nanostructures is realized if the covered and uncovered regions display different electronic properties which has only been demonstrated for the homogeneous systems[2], see Fig 1c. Scanning Tunneling Spectroscopy (STS) would seem to be the ideal tool to show if different gaps are also present in our nanostructures. Nevertheless, we were not able to obtain unambiguous data in order to detect noticeable changes in the evolution of the LDOS as the clusters were subsequently removed. In fact, for graphene on Ir(111) surfaces, dI/dV spectra seem to be mostly sensitive to a holelike surface resonance of the Ir(111) substrate rather than to any states of the graphene layer which was attributed to the selectivity of the tunneling current for states with small parallel momentum [41]. But even though transport and STS measurements are difficult due to the metallic substrate, optics and plasmonics seem within reach and in the following we discuss two new features that have the potential for sensors, metamaterials or data processing.

First, in the graphene/Ir(111) system, plasmonic excitations have been measured by electron energy loss spectroscopy[42]. We propose that they could be used to reach high field intensities since they are related to $\pi \to \pi^*$ transitions between the valence and conduction band, so-called interband plasmons[43]. Assuming local band-gap variations between covered and uncovered graphene regions[2], interband processes with transition energies $0.1 \text{eV} \leq E \leq 0.4$ eV should be forbidden in the covered, but allowed in the uncovered regions and can thus be localized to small graphene areas by removing the upper Ir-clusters, see Fig. 4a. In this way, quantum dots/wires can be designed at will with 2.5 nm precision by selectively removing metallic clusters. Quantum dots/wires with diameter/width $L$ posses normal modes corresponding to the wave number $q = n\pi/L$ of the interband plasmon ($n \in \mathbb{N}$), we thus expect large field enhancement due to resonant feedback effects which might be used, for instance, for spectroscopy on macro-molecules. For charge resonances at in-plane momentum $q \approx 0.03(0.05)\text{Å}^{-1}$ and energy $E \approx 0.25(0.375)$ eV, [42] the predicted field enhancement would occur for characteristic dot/wire dimensions of $L \approx 10.5(6.2)$ nm for $n = 1$ or $L \approx 21(12.5)$ nm for $n = 2$. These length scales are well within the reach of our technique (see supplementary information for details). In the same way, similar ideas can be applied to periodic structures where the excitation can be achieved also via propagating light.

A second and exciting new feature is given by the possibility to confine charged carriers, i.e., electrons as well as holes, within arbitrary geometrical regions due to locally modifying the electronic gap. One could hence design graphene quantum dots or nanowires of arbitrary size and form limited only by the cluster size of 2.5nm which has to be contrasted with graphene nanostructures obtained by electron beam lithography and subsequent etching which have typical dimensions L = 20 − 100nm. Using the effective-mass-approximation and thus the standard Dirac Hamiltonian with a variable mass profile, the discrete spectrum of a circular quantum dot as function of the radius $R$ can be obtained, see Fig 4c and the supplementary information. As indicated by the unshaded region, it displays only one localized state for $R \leq 7$ nm. In this regime, the uncovered area could resemble a quantum bit with qubit states "zero exciton" or "one exciton". The excitonic states can further arbitrarily be connected by conventional wave function overlap or via Förster energy transfer which is mediated by the Coulomb interaction between the excitonic states, see Fig 4d. This would lead to the emergence of excitonic bands with high lifetimes as estimated via Fermi's Golden Rule (see supplementary information).

The feasibility of the above proposals crucially depends on the impact that the Iridium substrate and the clusters on top have on the electronic properties of the graphene layer. Even though from ARPES experiments the band structure hardly seems to be affected beyond the gap of $\Delta \approx 0.1$eV and $\approx 0.4$eV, respectively, the graphene Dirac cone has been reported to hybridize near the Fermi level with the S1 surface state of Ir(111)[28], and also graphene's lattice structure changes from $sp^2$- to $sp^3$-bonding on the covered regions[33].

Additionally, graphene optics on a metallic substrate is challenging since the induced electric dipoles in the graphene layer are usually strongly quenched by the metallic substrate. Screening effects of the underlying Iridium acting as a metallic gate will further limit the lifetime of the electron-hole pairs[47]. The implications of the Ir-substrate involving optical (q = 0) transitions, and consequently the feasibility of the proposed emergence of excitonic bands, thus need to be tested experimentally. Nevertheless, as revealed by our analysis (see SI) on the experimentally measured plasmonic dispersion on graphene on Ir(111)[42], the screening influence of the metal on the charge excitations with finite q is surprisingly small suggesting that plasmonic excitations involving finite q-transitions should be almost unaffected by the Ir-substrate.

To conclude, we have presented a perfectly reproducible nanopatterning technique for graphene that combines bottom-up with top-down approaches. The precision is related to the periodicity of the Moiré-pattern that is formed by the graphene layer with the underlying substrate. Presupposing locally distinct electronic gaps in the covered and uncovered regions, new devices could be tailored with nano precision and we propose a novel platform for plasmonics relying on inter- rather than on intraband transitions. Also single graphene quantum dots/wires could be designed at will and arranged to arbitrary circuitries. Determining the optical gap and relaxation properties of mass-confined Dirac electrons via optical near-field scanning spectroscopy, emission spectroscopy or even transmission spectroscopy by chemically reduce the thickness of the sample would provide new insight on the role of the Ir(111) substrate on the excitonic decay rate. Finally, we note that the electronic spectrum drastically changes in the presence of a magnetic field due to the appearance of the zeroth Landau level not present in conventional semiconductor quantum dots which could be observable via Terahertz magneto-Raman spectroscopy.

**METHODS**.

The STM experiments were performed with a home-built variable temperature instrument [44,45]. Tips were made of W and prepared by electrochemically etching and subsequently annealing in UHV conditions. STM data were acquired with a fully automated workstation that incorporates digital feedback control based on DSP (digital signal processor) technology. All the surface manipulation experiments, data acquisition, and image processing were performed using the WSxM software [46]. STM images were all acquired in the constant current mode.

**Sample preparation:** Ir(111) surfaces were cleaned by 1 keV $Ar^+$ sputtering at 850 ºC. The growth of graphene on the clean Ir(111) surface was performed by chemical vapor deposition (CVD) of ethylene ($3\times10^{-7}$ Torr during 1 min) in UHV environments with the Ir(111) substrate held at 1050 ºC. Under such conditions, small areas of the Ir(111) substrate remained intentionally uncovered by graphene, which allowed us to estimate the coverage of W or Ir used for the cluster formation. W and Ir were evaporated from high purity filaments composed of each corresponding material. An accurate calibration of the deposition rate as a function of the filament temperature, measured by an infrared pyrometer, was performed by means of STM images acquired on areas of bare - uncovered by graphene- Ir(111).

**Theory and Modeling:** The experimental data (plasmonic excitations in graphene on Ir(111) and Pt(111)) was obtained from the original publications and fitted to the theoretical predictions using the least-square method. The electronic properties of graphene were modeled using the standard effective-mass-approximation. Exciton lifetimes and hopping amplitudes were estimated via Fermi's Golden Rule.

**FIGURES.**

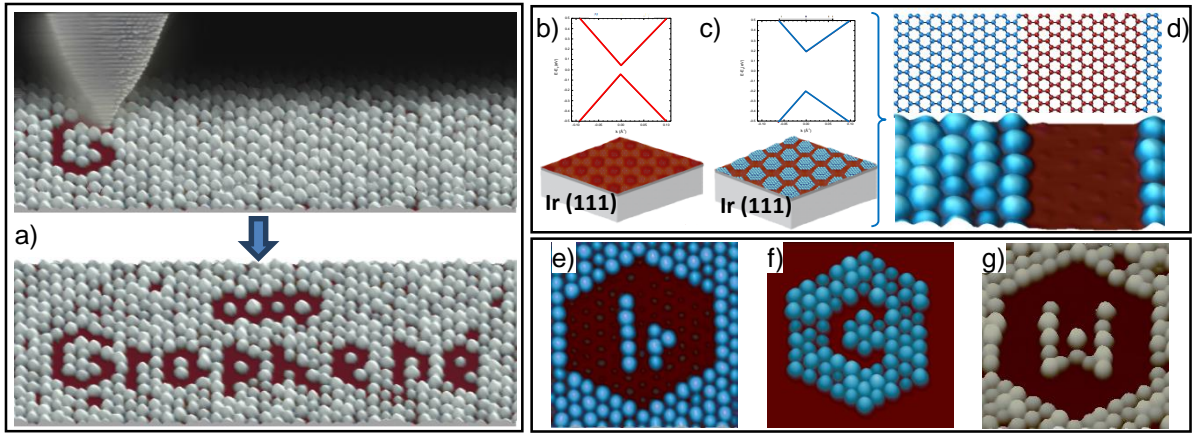

**Figure 1. Tailoring graphene with 2.5nm accuracy.** a) Upper panel illustrates the patterning process, with a schematic STM tip drawn on top of a real experimental image, removing selected W clusters from the G/Ir(111) surface to write the word "graphene". Lower panel shows a 95x35 nm$^2$ STM image with the final result. b), c) Graphene π bands, in the vicinity of $E_F$, for pristine G/Ir(111) and G/Ir(111) covered with an Ir cluster superlattice respectively, as measured by photoemission in ref (2). d) Example of a graphene-based nanostructure formed by two different "graphene" with the electronic properties depicted in b) and c). e)-g) 30x30 nm$^2$ STM images showing the validity of our method for clusters of different materials, in particular Ir (e,f) and W (g). Tunneling parameters: $I_T$ = 20 pA, $V_s$ = +2.2 V (a); $I_T$ = 150 pA, $V_s$ = +1.5 V (d); $I_T$ = 150 pA, $V_s$ = +1.5 V(e); $I_T$ = 160 pA, $V_s$ = +2.0 V (f); $I_T$ = 40 pA, $V_s$ = +1.5 V (g). We have used the following color code in all our images: reddish corresponds to pristine G/Ir(111), bluish to Ir clusters and grayish to W ones. All STM data were acquired and analyzed using the WSXM software [46]

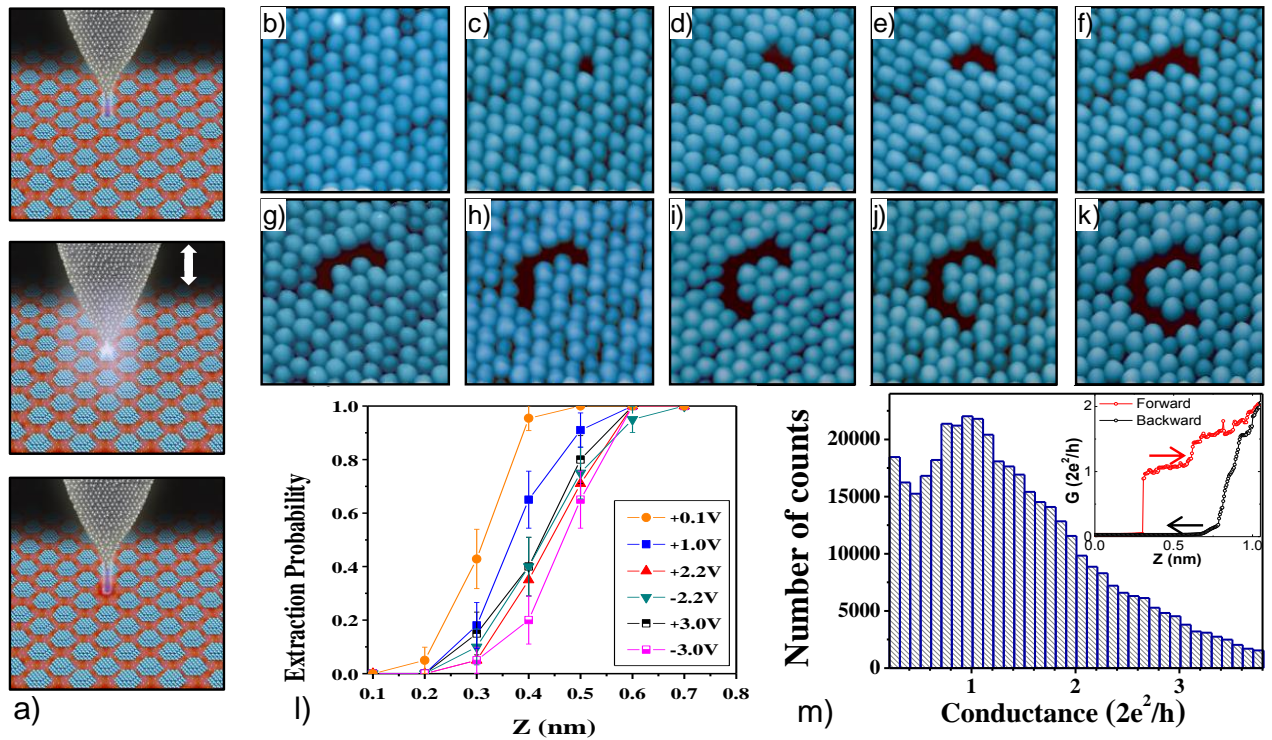

**Fig 2. Cluster extraction procedure.** a) Illustration of our cluster extraction method by the vertical displacement of a STM tip. b-k) Sequence of 23x23 nm$^2$ STM images, showing the writing of the carbon chemical symbol by consecutively removing one by one Ir clusters. All the images were acquired at RT with $I_T$ = 160 pA and $V_s$ = +2.0 V. To remove each cluster, the STM tip was approached 0.6 nm to the surface at 100 mV. l) Curves of the probability of removing a cluster as a function of approaching distance for several applied voltages. In all cases, stabilization current was set to 70 pA before opening the feedback loop. m) Conductance histogram constructed from I(z) measurements on 1200 single cluster extraction events. Each curve was obtained at RT, with 0.1V sample voltage. Inset shows an example of the conductance curve recorded during one of such cluster extraction events.

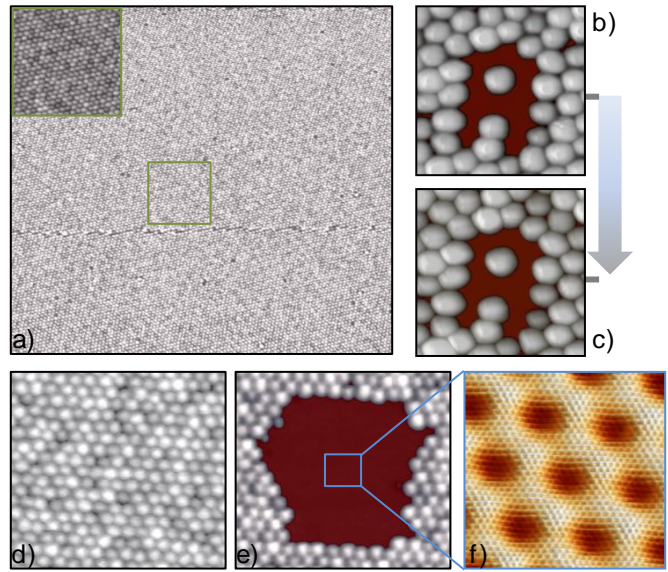

**Fig. 3. Nanostructures' size, stability and quality**. a) STM image of a 0.25x0.25μm² region fully covered by W clusters. Upper left inset shows a zoom of the region outlined by the green square. b,c) 16x16 nm² STM images of an artificially created nanostructure measured at room temperature with 24 hours difference. d) 40x40 nm² STM image of a W cluster superlattice on G/Ir(111). e) STM image showing the same region as in d) after deliberately removing a large number of clusters from it. f) 7.2x7.2 nm² STM image showing, with atomic resolution, the region outlined by a blue square in e). Tunneling parameters: $I_T$ = 50 pA, $V_s$ = 1.5 V (a); $I_T$ = 270 pA, $V_s$ = 2.3 (b, c); $I_T$ = 50 pA, $V_s$ = 2.2 V(d, e); $I_T$ = 5 nA, $V_s$ = 35 mV (f)

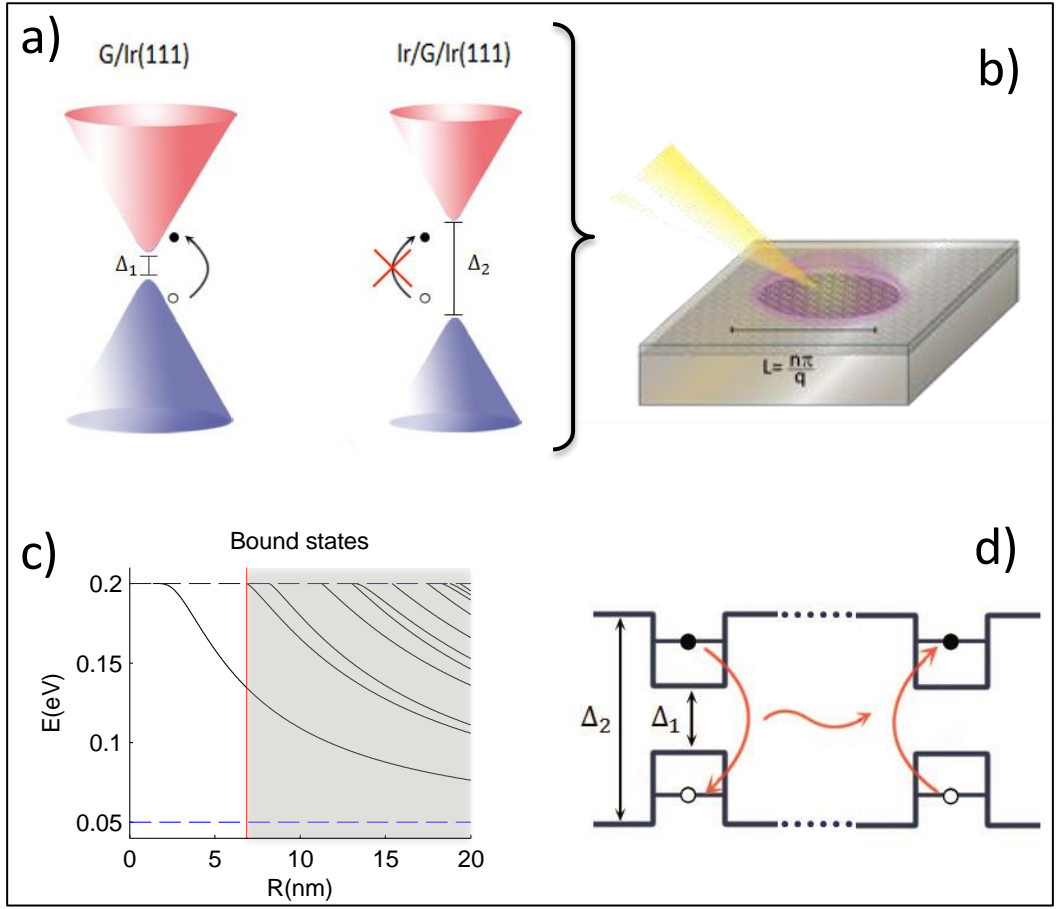

**Fig. 4: Field enhancement and mass confined quantum dots for Dirac electrons**. a) Energy dispersion near the Dirac points of the uncovered and covered areas displaying different mass gaps $\Delta_1 < \Delta_2$. Interband transitions with energies $\Delta_1 < \hbar\omega < \Delta_2$ are assumed to be allowed in the uncovered dot, but forbidden in the surrounding region covered with Iridium clusters. b) Schematic view of the set up leading to field enhancement by resonantly exciting interband acoustic plasmons by an electron beam or near-field techniques. c) Energy spectrum of a circular quantum dot with radius $R$ and discontinuous mass profile m= $\Delta_1\theta(R - r) + \Delta_2\theta(r - R)$ where the mass parameters are $\Delta_1 = 0.1$ eV and $\Delta_2 = 0.4$ eV. The unshaded region left from the red vertical line indicates the qubit regime of only one excitonic state. d) Schematic energy diagram of two mass confined quantum dots in the qubit regime. Electron-hole excitations (excitons) can efficiently change sites

via Förster energy transfer rate which is several orders of magnitude larger than the intrinsic decay rate.

**Supporting Information**.

Further experimental data with details about the sample preparation, the extraction probability and the conductance histogram are provided. Theoretical details discussing the acoustic interband plasmons in graphene on Iridium(111) and field enhancement, the spectrum of mass-confined graphene quantum dots and the excitonic bands in arbitrary quantum dot are also provided.

**Corresponding Author**

*Correspondence to: ivan.brihuega@uam.es.

Additional Information:

Author Contributions

A.J.M.G carried out the experiments supported by I.B. and J.M.G.R. A.J.M.G. and J.M.G.R. designed the research helped by I.B. A.J.M.G, I.B and J.M.G.R. analyzed the data. T.S performed the main calculations with the collaboration of A.G.R. I.B wrote the manuscript with the help of T.S. All authors contributed to the scientific discussion and revised the manuscript.

**Competing financial interests**

The authors declare no competing financial interests

Funding Sources

This work was supported by Spain's MINECO under Grants No. MAT2013-41636-P, No. FIS2013-44098, No. CSD2010-00024; by Comunidad de Madrid under Grant No. CPI/0256/2007.

Notes

Any additional relevant notes should be placed here.

**Acknowledgment**

We thank F. Guinea, G. Gómez-Santos and B. Amorim for helpful discussions.

# Supplementary Material for the article "Towards scalable nano-engineering of graphene"

A.J. Martínez-Galera[1], I. Brihuega[1,2,*], A. Gutiérrez-Rubio[1,3],
T. Stauber[1,2,3], and J. M. Gómez-Rodríguez[1,2]

[1] *Departamento de Física de la Materia Condensada,
Universidad Autónoma de Madrid, E-28049 Madrid, Spain*

[2] *Departamento de Física de la Materia Condensada,
Universidad Autónoma de Madrid, E-28049 Madrid, Spain*

[3] *Instituto de Ciencia de Materiales de Madrid,
Consejo Superior de Investigaciones Cientficas, E-28049 Madrid, Spain*

*) Corresponding author: ivan.brihuega@uam.es

# EXPERIMENTAL DETAILS

## Sample preparation

The experiments were carried out in an ultra-high-vacuum (UHV) system whose base pressure is below $5 \times 10^{-11}$ Torr. The system is equipped with a variable temperature scanning tunneling microscope (VT-STM), low energy electron diffraction (LEED), Auger electron spectroscopy (AES), sample and STM tip transfer and heating capabilities, an STM tip cleaning system by field emission, several interchangeable evaporation cells, a quartz crystal microbalance, and an ion gun for sample cleaning purposes.

The STM experiments were performed with a home-built variable temperature instrument. Tips were made of W and prepared by electrochemically etching and subsequently annealing in UHV conditions. STM data were acquired with a fully automated workstation that incorporates digital feedback control based on DSP (digital signal processor) technology[1]. All the surface manipulation experiments, data acquisition, and image processing were performed using the WSxM software[2]. STM images were all acquired in the constant current mode.

Ir(111) surfaces were cleaned by 1 keV Ar$^+$ sputtering at 850 °C. The growth of graphene on the clean Ir(111) surface was performed by chemical vapor deposition (CVD) of ethylene ($3 \times 10^{-7}$ Torr during 1 min) in UHV environments with the Ir(111) substrate held at 1050 °C. Under such conditions, small areas of the Ir(111) substrate remained intentionally uncovered by graphene, which allowed us to estimate the coverage of W or Ir used for the cluster formation (see below). W and Ir were evaporated from high purity filaments composed of each corresponding material. An accurate calibration of the deposition rate as a function of the filament temperature, measured by an infrared pyrometer, was performed by means of STM images acquired on areas of bare -uncovered by graphene- Ir(111). Figure S1 shows a representative STM topograph acquired after the adsorption of $0.50 \pm 0.05$ ML of W at a deposition rate of $8 \times 10^{-3}$ ML/s on a graphene/Ir(111) surface at room temperature. Here, 1 ML corresponds to the atomic density of the Ir(111) surface. As can be appreciated in Fig. S1, a W cluster superlattice with high structural perfection extending over very large areas of the graphene/Ir(111) surface is formed.

### Extraction probability

We acquired more than 1000 curves in order to obtain the extraction probability shown in Fig. 2l of the main manuscript. Each single point in this graph corresponds to the probability of removing a W cluster by approaching the STM tip towards the sample a certain distance with a given applied sample voltage. For each point, this probability was obtained by performing 20 single attempts to extract 20 different clusters. In each attempt, we set the stabilization current to 70 pA at the selected sample voltage and then, with the selected voltage fixed, we approached the tip towards the surface the chosen distance at a constant rate. After each single attempt, we measured a STM image to check whether the selected cluster was removed or not. The extraction probability for each point of the graph corresponds then to the total number of clusters removed in these 20 attempts divided by 20. Error bars correspond to the standard deviation of a binomial distribution.

### Conductance histogram

We constructed the conductance histogram shown in Fig. 2m of the main manuscript by recording the current during the vertical displacement of the STM tip (I-Z curves) on 1200 single W cluster extraction events. Each I-Z curve was obtained, at RT, by approaching 1.0 nm the STM tip towards the sample, with 0.1 V sample voltage and stabilization current of 70 pA before opening the feedback loop. In all cases we verified by STM that the metallic cluster was successfully removed. Such conductance histogram was required to extract some meaningful information from the experimental data since, as explained in the main text, individual conductance curves so acquired were inherently irreproducible. In Fig. S2, we present four typical I-Z curves, each of them acquired during the extraction of a W cluster, to illustrate this point.

# THEORETICAL DETAILS

**Acoustic interband plasmons in graphene on Iridium(111) and field enhancement**

*In this section, we will propose that experimentally observed plasmonic resonances in graphene on Ir(111) [T. Langer et al., New J. Phys. **13**, 053006 (2011)] are due to interband transitions and that the screening influence of the metal on the charge excitations with finite wave number is surprisingly small.*

In Ref. 3, graphene on Iridium(111) was investigated by means of high resolution electron energy-loss spectroscopy (EELS). A characteristic peak in the loss function with linear dispersion from energies $\hbar\omega = 0.1\text{eV}$ up to $\hbar\omega = 1.5\text{eV}$ was seen. Additionally, the width of the resonance showed linear behavior. Both features can be captured by a simple theory assuming the conductivity for undoped graphene including only interband transitions,[4]

$$\sigma(q,\omega) = \sigma_0 \frac{\omega}{\sqrt{\omega^2 - (v_F q)^2}}, \qquad (1)$$

with $\sigma_0 = e^2/(4\hbar)$ the universal conductivity and the Fermi velocity $v_F$. With the dielectric function

$$\epsilon(q,\omega) = 1 + \frac{i\sigma(q,\omega)q}{2\omega\epsilon\varepsilon_0}, \qquad (2)$$

and the (generally complex) static dielectric constant $\epsilon = \epsilon' + i\epsilon''$, the energy loss function $S = -\text{Im}\,\epsilon^{-1}$ shows a maximum at $x = 1$ with

$$x = \frac{\pi\alpha_g}{2|\epsilon|} \frac{v_F q}{\sqrt{\omega^2 - (v_F q)^2}}, \qquad (3)$$

$|\epsilon| = \sqrt{\epsilon'^2 + \epsilon''^2}$ and graphene's fine-structure constant $\alpha_g = \alpha \frac{c}{v_F} \approx 2.2$ where $\alpha = e^2/(4\pi\varepsilon_0 \hbar c) \approx 1/137$. This corresponds to charge excitations with a linear dispersion $\omega = v_s q$ where the sound-velocity reads $v_s = \sqrt{1 + \left(\frac{\pi\alpha_g}{2|\epsilon|}\right)^2}\, v_F$. For $|\epsilon| \approx 3.5$, we obtain a good fit to the experimentally observed sound velocity of $v_s \approx 1.4 v_F$ for Iridium.[3]

EELS experiments were also performed for graphene on SiC[5] and on Platinum.[6] Again, a characteristic peak in the loss function with linear dispersion was observed with sound velocities $v_s \approx 1.4 v_F$ for a SiC-substrate and $v_s \approx 1.15 v_F$ for a Platinum substrate. Applying the same theory as above, i.e., only including interband transitions, we obtain good fits to the experimental data for $|\epsilon| \approx 3.5$ (SiC) and $|\epsilon| \approx 6.1$ (Platinum), respectively. For a dielectric substrate, the dielectric response is real, $|\epsilon| \approx \epsilon'$, leading to the usual dielectric constant for SiC $\epsilon_{SiC} \approx 6$, where we used $\epsilon' = (\epsilon_{SiC}+1)/2$ valid for a substrate/graphene/air interface. This confirms that assuming interband plasmonic resonances leads to consistent results.

We can refine the above approach and also include the band gap $\Delta$ that appears in the spectrum of graphene on Iridium(111). For that, we need to consider the conductivity of gapped graphene which also acquires an imaginary part. The longitudinal part that couples to charge fluctuations reads[7]

$$\text{Re}\{\sigma(q,\omega)\} = \sigma_0 \sqrt{1-\left(\frac{v_F q}{\omega}\right)^2}^{-1} \left[1+\frac{\Delta^2}{\hbar^2(\omega^2-(v_F q)^2)}\right] \theta\left((\hbar\omega)^2-(\hbar v_F q)^2-\Delta^2\right); \qquad (4)$$

$$\text{Im}\{\sigma(q,\omega)\} = \frac{\sigma_0}{\pi} \Bigg\{ \frac{2\Delta}{\hbar\omega}\frac{\omega^2}{\omega^2-(v_F q)^2} + \frac{2\omega}{\sqrt{|(v_F q)^2-\omega^2|}}\left[1+\frac{\Delta^2}{\hbar^2(\omega^2-(v_F q)^2)}\right] \times \qquad (5)$$

$$\times \left[\theta(v_F q-\omega)\arccos\left(\frac{\Delta}{\sqrt{\Delta^2+\hbar^2((v_F q)^2-\omega^2)}}\right) - \theta(\omega-v_F q)\frac{1}{2}\log\left|\frac{\hbar\sqrt{\omega^2-(v_F q)^2}+\Delta}{\hbar\sqrt{\omega^2-(v_F q)^2}-\Delta}\right|\right] \Bigg\}.$$

On the left hand side of Fig. S3, the electron loss function $S = -\text{Im}\,\epsilon^{-1}$ with the conductivity of gapped graphene, Eqs. (4) and (5), is shown. We also plot the experimental data of Ref. 3 (squares) and the linear dispersion $\omega = v_s q$ with sound velocity $v_s = \sqrt{1+\left(\frac{\pi\alpha_g}{2|\epsilon|}\right)^2}\, v_F$ for $|\epsilon| \approx 3.5$ (solid line). There are no significant changes compared to the theory of ungapped graphene. This suggests that the charge resonances seen in typical EELS experiments are due to interband transitions which can be confined up to a certain energy by a larger gap present in the surrounding region. An analogous analysis for gapless but doped graphene on Pt is carried out on the right hand side of Fig. S3, again yielding a good fit for $|\epsilon| \approx 6.1$. Let us finally note that both dielectric constants for Iridium as well as for Platinum are considerably lower than expected for a typical metal which suggests that optics for frequencies ranging from THz to mid-infrared associated with finite wave numbers is possible even in the presence of a metallic substrate.

### Spectrum of mass-confined graphene quantum dots

*In this section, we discuss the spectrum of spherical graphene quantum dots. Electronic localization is obtained due to a different mass-profile presumably provoked by the presence or absence of the metallic nanoclusters.*

The removal of Ir-clusters on top of graphene on Iridium(111) presumably creates Dirac quantum dots with nanoscale dimensions due to mass confinement. Dirac carriers localized by a variable mass profile lead to well-defined discrete states with high intrinsic lifetime due to the absence of typical edge disorder. We model the confined region by a circular quantum dot of radius $R$ where a step-like change in the mass term $2mv_F^2 = \Delta_1 \theta(R-r) + \Delta_2 \theta(r-R)$ with $\Delta_1 = 0.05\text{eV}$ for the dot and $\Delta_2 = 0.4\text{eV}$ for the bulk region. The Hamiltonian of a gapped graphene sheet in polar coordinates reads

$$\hat{H} = \begin{bmatrix} mv_F^2 & -i\hbar v_F e^{-i\theta}\left[\partial_r - \frac{i}{r}\partial_\theta\right] \\ -i\hbar v_F e^{i\theta}\left[\partial_r + \frac{i}{r}\partial_\theta\right] & -mv_F^2 \end{bmatrix}. \tag{6}$$

With $l \in \mathbb{Z}$, the eigenfunctions can be written in the general form

$$\psi_{E,l}(\vec{r}) = \langle \vec{r}|El\rangle = \begin{bmatrix} f_{E,l}(r)e^{il\theta} \\ g_{E,l}(r)e^{i(l+1)\theta} \end{bmatrix}. \tag{7}$$

For the given mass profile and $\Delta_1 < 2|E| < \Delta_2$, we explicitly have

$$f_{E,l}(r) = \begin{cases} B j_l(qr) & \text{for } r < R \\ BA_l h_l^{(1)}(kr) & \text{for } r > R \end{cases}, \tag{8}$$

$$g_{E,l}(r) = \begin{cases} Bi\eta_q j_{l+1}(qr) & \text{for } r < R \\ BA_l i\eta_k h_{l+1}^{(1)}(kr) & \text{for } r > R \end{cases}, \tag{9}$$

with $j_l$ the Bessel and $h_l^{(1)} = j_l + iy_l$ the Hankel function where $y_l$ denotes the Neumann function. We further have with $\Delta_1 = 2m_1 v_F^2$ and $\Delta_2 = 2m_2 v_F^2$

$$A_l = \frac{j_l(qR)}{h_l^{(1)}(kR)}, \tag{10}$$

$$q = \sqrt{(E - m_1 v_F^2)(E + m_1 v_F^2)}/\hbar v_F, \qquad \eta_q = \text{sg}(E)\sqrt{\left|\frac{E - m_1 v_F^2}{E + m_1 v_F^2}\right|}, \tag{11}$$

$$k = i\sqrt{|(E - m_2 v_F^2)(E + m_2 v_F^2)|}/\hbar v_F, \qquad \eta_k = i\sqrt{\left|\frac{E - m_2 v_F^2}{E + m_2 v_F^2}\right|}, \tag{12}$$

and $B$ is a normalization constant. The matching conditions on the frontier $R$ yield the

following equation:

$$\eta_k h^{(1)}_{l+1}(kR) j_l(qR) = \eta_q j_{l+1}(qR) h^{(1)}_l(kR) \ . \qquad (13)$$

Its solutions are given by the discrete bound state energies. These are shown in Fig. S4, where the black curves correspond to $l \geq 0$ and the red curves to $l < 0$. On the left hand side of Fig. S5, the eigenfunctions of Eq. (8) and (9) are plotted for the first three electronic energy levels of a dot with radius $R = 15$nm and on the right hand side the probability density $|\psi_{E,l}(\vec{r})|^2$.

There is a bound state at $E \approx 0.2$eV for a radial quantum dot even with radius $R = 2.5$nm. This energy is lowered by excitonic effects and also due to a smooth effective mass evolution which tends to decrease the electron confinement. Changes might also occur due to strain effects leading to effective magnetic fields.[8] We thus predict a bound exciton even in small quantum dots with radius $R = 2.5$nm to be verified by photoluminescence or near-field spectroscopy. For larger, not necessarily circular quantum dots, we also expect at least one bound state in the conduction as well as in the valence band, see Fig. S4. For quantum dots with more than one bound state, the relaxation mechanisms to the ground state due to various mechanisms such as optical/acoustic phonons or Auger electrons can be discussed in detail as the dot size can be increased till the two-dimensional limit has been reached.

## Excitonic bands in arbitrary quantum dot arrays

*In this section, we will discuss possible extensions and future research directions that our approach offers. For explicit calculations, we assume a non-dissipative substrate which needs to be improved in subsequent calculations after the screening influence of the metal has been experimentally investigated.*

Graphene optics on a metallic substrate is challenging since the induced electric dipoles in the graphene layer are usually strongly quenched by the metallic substrate. Nevertheless, the previous analysis on the acoustic plasmonic excitations showed that the influence for finite $q$-numbers is considerably less than expected, especially for Iridium, leading to a static dielectric constant $|\epsilon| \approx 3.5$. This means that screening due to induced particle-hole processes in Iridium is suppressed and that the lifetime of graphene's electron-hole excitations is only weakly limited by the substrate. The suppressed fluorescent quenching in the relevant (optical) regime would open up the possibility for optical experiments on

graphene/Ir(111) at THz to mid/near infrared frequencies. This holds even more if finite wave numbers $q$ are involved in the absorption process which can be the case in periodic superstructures.

In the following, we will thus neglect intrinsic damping and assume a *real* dielectric constant $\epsilon$. Then, only the spontaneous decay of the excitons must be considered when evaluating their lifetimes. This decay rate is given by[9]

$$\gamma = \frac{\omega^3 |\vec{\mu}|^2}{3\pi\varepsilon_0 \epsilon \hbar c^3}, \quad (14)$$

where $\omega$ and $\vec{\mu}$ are the energy and the dipole moment of the exciton, respectively, $\varepsilon_0$ the vacuum permittivity and $\epsilon$ the real (average) dielectric constant, i.e., $\epsilon = (\epsilon_{substrate} + 1)/2$ for a substrate/graphene/air interface. For graphene, we can write this as

$$\gamma = \frac{4\alpha\omega}{3\epsilon}\left(\frac{v_F}{c}\right)^2 |\langle El|\sigma_x|E'l'\rangle|^2 \quad (15)$$

with $|El\rangle$ ($|E'l'\rangle$) the electron (hole) bound state, $\hbar\omega = E - E'$, $\alpha$ the fine-structure and $\sigma_x$ the $x$-component of the Pauli-matrices. In a setup with only one bound state in the valence and conduction band, like the one depicted in Fig. S6, one has $E = -E' = \hbar\omega/2$ and $l = 0, l' = -1$. For $R = 5$nm and $\epsilon \approx 3.5$, this yields $\gamma \simeq 2 \cdot 10^7 \text{s}^{-1}$ at transition energy $\hbar\omega \approx 0.3$eV.

Let us now assume a quantum dot in the first excited (excitonic) state next to a second quantum dot at distance $D$ of the same dimension that is in its ground state. Since they have the same excitation spectrum there is the possibility for efficient energy (Förster) transfer from one dot to the other, see process 2 in Fig. S6. If this non-radiative process is considerably faster than the spontaneous decay rate of the exciton, a hopping mechanism for the exciton is induced and arbitrary two-dimensional lattices could be designed with the life-time limited only by intrinsic dissipative effects like phonon and bulk-electron scattering.

The transition probability $t_n$ of an exciton hopping from one quantum dot to another mediated by the inter-dot Coulomb interaction $V_C$ shall be estimated using Fermi's Golden rule,

$$t_n = \frac{2\pi}{\hbar}\sum_m |\langle n|V_C|m\rangle|^2 \delta(\epsilon_n - \epsilon_m), \quad (16)$$

where $|m\rangle$ and $|n\rangle$ are excitonic states. The delta function is usually replaced by the absorption and emission spectra of the donor and acceptor. We will thus assume homogeneously broadened excitonic states $\rho_n(\epsilon) = \pi^{-1}\Gamma/((\epsilon - \epsilon_n)^2 + \Gamma^2)$ and replace the $\delta$-function by $\delta_{n,m} = \int d\epsilon \rho_n(\epsilon)\rho_m(\epsilon)$.[10] Conventional overlap of the wave functions of quantum dots leads to an additional hopping mechanism.

We model the excitonic state as product state of the electron and hole wave function. The matrix element is thus given by $\langle E l \vec{r}_2; E'l'\vec{r}_1|V_C|El\vec{r}_1; E'l'\vec{r}_2\rangle$, where $E' < 0$, $E > 0$ and $V_C$ the Coulomb interaction with

$$\langle \vec{r}_3 s_3; \vec{r}_4 s_4 | V_C | \vec{r}_1 s_1; \vec{r}_2 s_2 \rangle = \frac{e^2}{4\pi\varepsilon_0\epsilon|\vec{r}_1 - \vec{r}_2|}\delta_{s_1,s_3}\delta_{s_2,s_4}\delta(\vec{r}_1 - \vec{r}_3)\delta(\vec{r}_2 - \vec{r}_4) \ . \tag{17}$$

The two (direct and exchange) processes are depicted in Fig. S6.

Let us consider the special case of only one excitonic bound state, i.e., quantum dots with radius $R \lesssim 10$nm and we drop the subindex $t_n \to t$. For typical broading $\Gamma = 10$meV and dot radius $R = 5$nm, the hopping probabilities for the two non-radiative processes, depicted in Fig. S6, as well as for conventional wave function overlap are plotted in Fig. S7 as a function of the inter-dot distance $D$.

The Förster energy transfer is dominant for $D > 6R$ displaying the typical algebraic decay $t \sim D^{-6}$ obtained from a multipolar expansion. In this regime, the excitonic hopping induced by this mechanism is two orders of magnitude greater than the spontaneous decay rate $\gamma$. As a consequence, a well-defined band of Frenkel excitons is expected with Wannier functions well approximated by the individual quantum dot wave functions. On the other hand, for $D < 6R$, Wannier functions describing the exciton will be more spread and the resultant excitons would be more delocalized. A different approach including the array of quantum dots from the beginning is then necessary.

To conclude, our results demonstrate that depending on the dimensions of the quantum dots and the distance between them, we can engineer lattices hosting different kinds of excitons (Frenkel vs. Wannier) and with tunable excitonic hoppings between the dots. This would provide an alternative platform to study the dynamics in quantum mechanical tight-binding models and also the possibility of discussing exciton-polariton condensation[11] in a controllable system.

## Acknowledgments

We thank F. Guinea, G. Gómez-Santos and B. Amorim for helpful discussions. This work has been supported by FCT under grants PTDC/FIS/101434/2008; PTDC/FIS/113199/2009 and MIC under grant FIS2010-21883-C02-02.

---

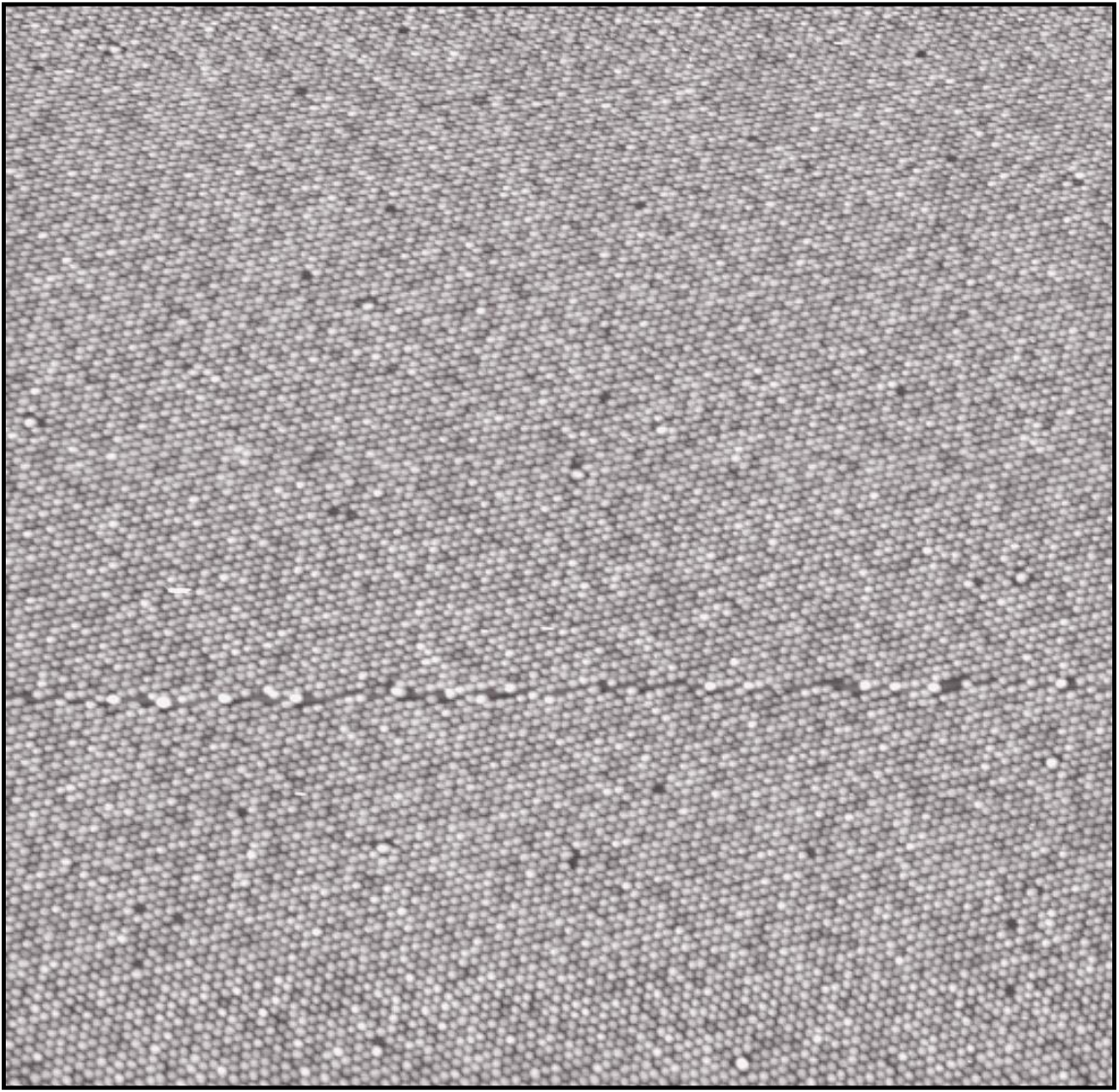

Fig. S1. Sample morphology. STM image of a 0.25x0.25μm$^2$ region fully covered by W clusters. Sample voltage: 1.5V; tunneling current: 50 pA

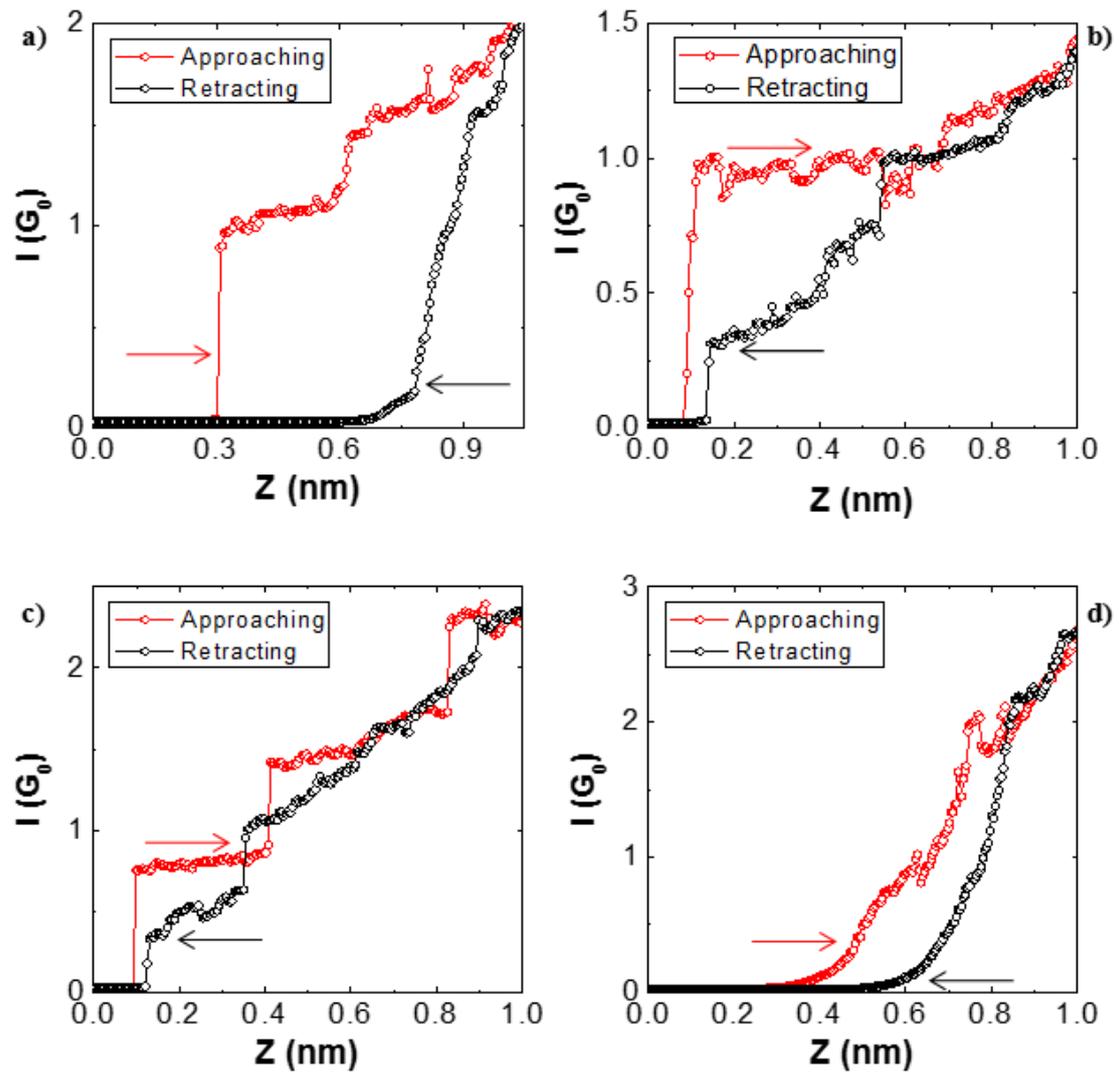

Fig S2. Characteristic I-Z curves illustrating the different behavior of the conductance traces in the cluster extraction process. All the curves were measured at RT with the same tunneling conditions; sample voltage =+100mV, stabilization current =70pA.

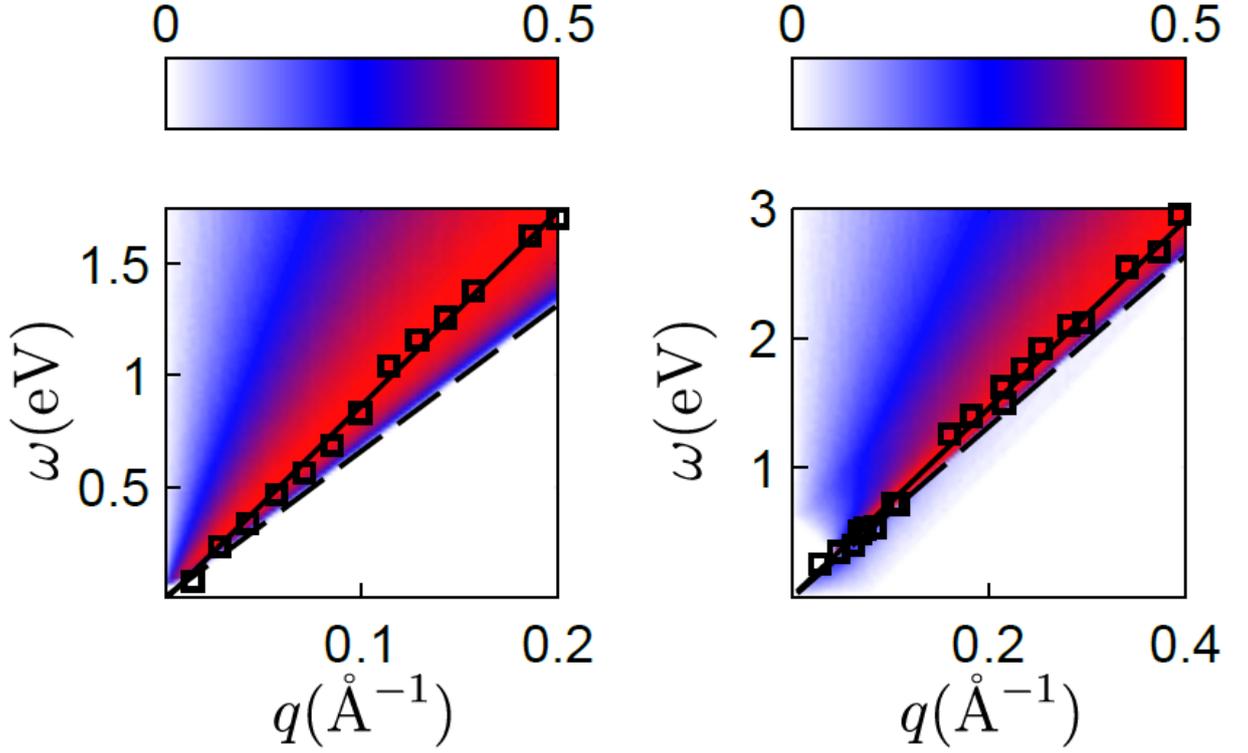

Fig. S3: Loss function of graphene on top of an Iridium (left) and Platinum (right) substrate due to interband transitions compared to the experimental data of Ref. 3 and 6 (squares), respectively. Also shown the acoustic plasmon dispersion $\omega = v_s q$ with sound velocity $v_s \approx \sqrt{1 + (\frac{\pi \alpha_g}{2|\epsilon|})^2} v_F$ for dielectric constants $|\epsilon| = 3.5$ (left) and $|\epsilon| = 6.1$ (right). The dashed line indicates the Dirac dispersion $\omega = v_F q$.

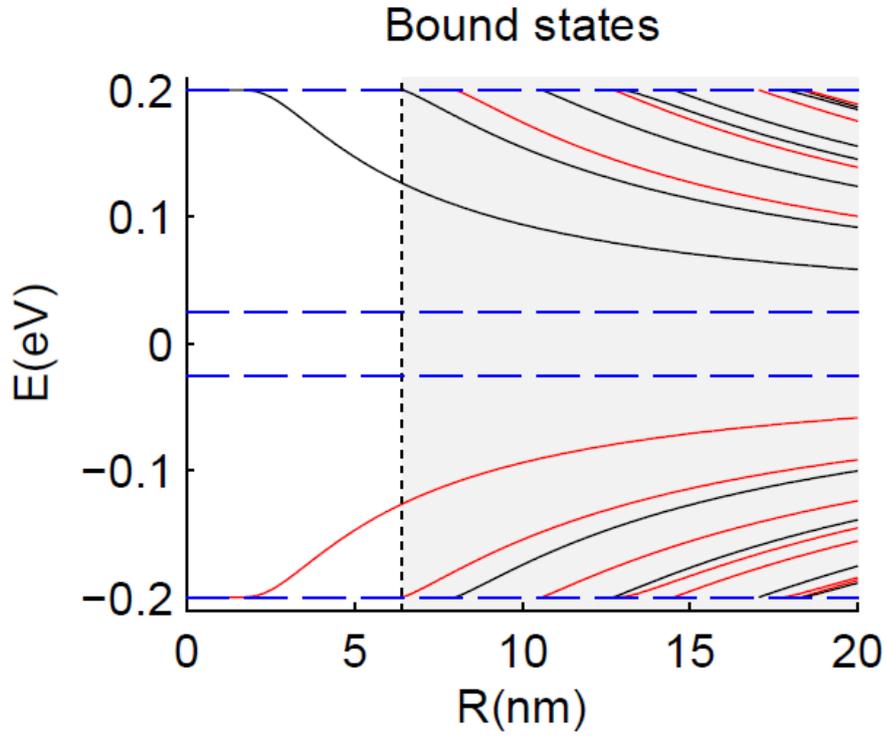

Fig. S4: Band structure of a circular graphene quantum dot with step-like mass profile $nv_F^2 = \Delta_1\theta(R-r) + \Delta_2\theta(r-R)$ with $\Delta_1 = 0.05\text{eV}$ for the dot and $\Delta_2 = 0.4\text{eV}$ for the bulk region. Black curves correspond to $l \geq 0$ and red curves to $l < 0$. To every curve with $l$ corresponds a particle hole symmetric curve with $-l-1$.

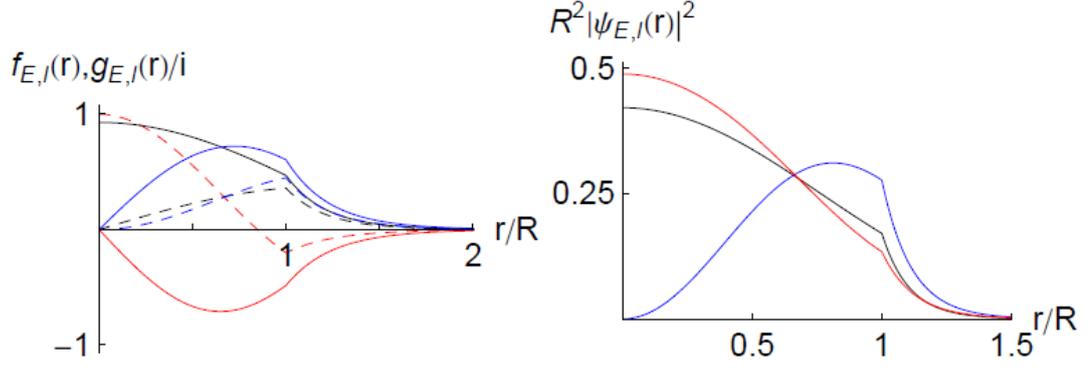

Fig. S5: Left hand side: $f_{E,l}(r)$ (solid lines) and $g_{E,l}(r)/i$ (dashed lines) in arbitrary units for a quantum dot with radius $R = 15\,\text{nm}$. The three lowest bound states with positive energy are shown in increasing order as black (lowest energy), blue and red curves which correspond to $l = 0$, $l = 1$ and $l = -1$, respectively. Right hand side: Probability density of the aforementioned bound states with $|\psi_{E,l}(r)|^2 = |f_{E,l}(r)|^2 + |g_{E,l}(r)|^2$.

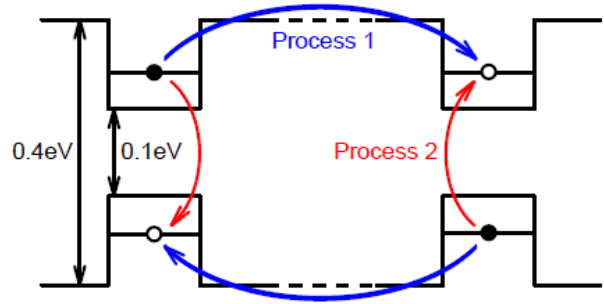

Fig. S6: Schematic band structure of the graphene quantum dot localized by a mass-term and the energy transfer from one dot to another via different processes mediated by the Coulomb interaction. Förster energy transfer corresponds to process 2.

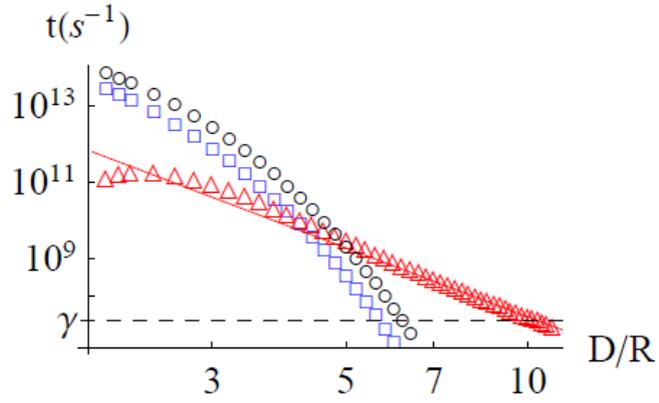

Fig. S7: (color online): Transition rates of different hopping processes between two quantum dots as function of the distance $D$ (see Fig. S6). Blue squares: process 1; red triangles: process 2; black circles: wave functions overlap. The solid red line corresponds to the multipolar expansion of the Coulomb potential and shows the typical algebraic Förster decay $t \sim D^{-6}$.